\documentclass[10pt]{article}
\usepackage[OE]{express}
\usepackage{mathtools}
\usepackage{ulem}
\DeclarePairedDelimiter\bra{\langle}{\rvert}
\DeclarePairedDelimiter\ket{\lvert}{\rangle}
\DeclarePairedDelimiterX\braket[2]{\langle}{\rangle}{#1 \delimsize\vert #2}
\DeclarePairedDelimiter\expect{\langle}{\rangle}

\usepackage{xcolor}
\begin{document}

\title{Hong-Ou-Mandel Interference of Unconventional Temporal Laser Modes}

\author{{\color{gray}Sascha Agne\authormark{*}\authormark{1,2}, Jeongwan Jin\authormark{1,3}, Katanya B. Kuntz\authormark{1}, Filippo M. Miatto\authormark{1,4}, Jean-Philippe Bourgoin\authormark{1,5}, and Thomas Jennewein\authormark{\S}\authormark{1,6}}}

\address{\authormark{1}Institute for Quantum Computing, Department of Physics and Astronomy, University of Waterloo, Waterloo, Ontario N2L 3G1, Canada}
\address{\authormark{2}Present Address: Max-Planck Institute for the Science of Light, Staudtstrasse 2, 91058 Erlangen, Germany}
\address{\authormark{3}Present Address: National Research Council of Canada, 1200 Montreal Road, Ottawa, ON K1A 0R6, Canada}
\address{\authormark{4}Present Address: Institut Polytechnique de Paris and T\'{e}l\'{e}com Paris, LTCI, 19 Place Marguerite Perey, Palaiseau 91120, France}
\address{\authormark{5}Present Address: Aegis Quantum, Waterloo, ON, Canada}
\address{\authormark{6}Quantum Information Science Program, Canadian Institute for Advanced Research, Toronto, Ontario M5G 1Z8, Canada}

\email{\authormark{*}sascha.agne@mpl.mpg.de}
\email{\authormark{\S}thomas.jennewein@uwaterloo.ca}

\email{\authormark{*}sascha.agne@mpl.mpg.de}
\email{\authormark{\S}thomas.jennewein@uwaterloo.ca} 


\begin{abstract}
The Hong-Ou-Mandel (HOM) effect ranks among the most notable quantum interference phenomena, and is central to many applications in quantum technologies. The fundamental effect appears when two independent and indistinguishable photons are superimposed on a beam splitter, which achieves a complete suppression of coincidences between the two output ports. Much less studied, however, is when the fields share coherence (continuous-wave lasers) or mode envelope properties (pulsed lasers). In this case, we expect the existence of two distinct and concurrent HOM interference regimes: the traditional HOM dip on the coherence length time scale, and a structured HOM interference pattern on the pulse length scale. We develop a theoretical framework that describes HOM interference for laser fields having arbitrary temporal waveforms and only partial overlap in time. We observe structured HOM interference from a continuous-wave laser via fast polarization modulation and time-resolved single photon detection fast enough to resolve these structured HOM dips.
\end{abstract}

\section{Introduction}
Interference of light from two sources rests on the superposition principle of electromagnetic waves, and remains valid even when the sources emit light independently. If we consider electromagnetic waves from two independent optical emitters without a phase relationship, any first-order interference washes out when detector electronics are too slow to trace rapidly fluctuating electric fields \cite{Paul1986}. However, correlation measurements can recover interference effects, in particular when time-resolved single photon detectors are used \cite{Magyar1963,Pfleegor1967,Ou1988,Ou1989}.

Single photons in Fock states are naturally phase-independent and mixing them on an optical beam splitter results in a bunching effect, whereby detectors in the two output ports of a beam splitter register fewer coincidence detections than a fair Bernoulli process would allow. This effect of genuine two-photon interference \cite{Pan2012} was first observed by Hong, Ou, and Mandel (HOM) \cite{Hong1987}. HOM interference is observed by varying the optical delay between the two input photons, resulting in an anti-correlation signal whose shape depends on the photons' spectra, or equivalently, temporal waveforms.

The HOM interference effect is not restricted to single photon states, and has been observed with both thermal \cite{Chen2011} and laser light \cite{Kim2014,Liu2015,Liu2016}. It is counter-intuitive that two independent continuous-wave (CW) lasers can show a HOM interference pattern (at a reduced visibility). However, this pattern is observed by correlating the detection times of the two outputs, and the dip width depends on the coherence lengths of the two input fields \cite{Kim2014}. Though the quantum optical photon picture is frequently employed to explain the reduction in detected coincidences, the effect also has a semi-classical explanation \cite{Shapiro2012}, allowing for new optical measurement techniques \cite{Lebreton2013a,Lebreton2013b}.

We build on the theoretical and experimental work that has been conducted with single photons in long temporal wavepackets \cite{Legero2004}. The result is a theoretical framework that describes HOM interference of arbitrarily-shaped pulses with long coherence length and partial overlap in time. We then use well-defined pulsed coherent states that have a long coherence time ($2\:\mu s$) to explore whether it is possible to recover full HOM visibility between partially overlapped pulses. By using a polarization-modulated CW laser to generate $2.83$\,ns square wave shaped pulses, we create time-dependent wavepackets that have distinguishable polarization states. We can resolve structured HOM dips by adjusting the time delay between the two modulated inputs until there is a region of overlap where the polarization states are indistinguishable. Since the time resolution of our detectors (<$0.1$\,ns) is significantly shorter than the pulse duration, we can recover the full HOM visibility within the region of temporal overlap for both triangle and square wave HOM patterns.   
\section{Theoretical framework}
\label{Sec:Theory}
We first derive the standard HOM dip theory, apply it to coherent states, and then study the effect of polarization modulation on a timescale much shorter than the coherence time. Consider two input fields $\hat{E}_1^-(t)=\zeta_1(t)\hat{a}_1^{\dagger}(t)$ and $\hat{E}_2^-(t)=\zeta_2(t)\hat{a}_2^{\dagger}(t)$ to a symmetric beam splitter. Here, $\hat{a}_i^{\dagger}(t)$ and $\zeta_i(t)$ are, respectively, the time-dependent creation operators and envelope functions for optical modes $i=1,2$. In these expressions, we assume a narrow-band approximation and omit overall constants. Hence, the beam splitter output fields are given by
\begin{equation}
\begin{aligned}
\hat{E}_3^-(t)&=\frac{1}{\sqrt{2}}\bigg(\zeta_1(t)\hat{a}_1^{\dagger}(t)+\zeta_2(t)\hat{a}_2^{\dagger}(t)\bigg)\\
\hat{E}_4^-(t)&=\frac{1}{\sqrt{2}}\bigg(\zeta_1(t)\hat{a}_1^{\dagger}(t)-\zeta_2(t)\hat{a}_2^{\dagger}(t)\bigg)\,.\label{eq:outputfields}
\end{aligned}
\end{equation}
The two-photon coincidence rates measured in HOM interference are described by the second-order cross-correlation function
\begin{equation}
G^{(2x)}(t_3,t_4)\coloneqq\left\langle{\hat{E}_3^-(t_3)\hat{E}_4^-(t_4)\hat{E}_4^+(t_4)\hat{E}_3^+(t_3)}\right\rangle\,,\label{eq:secondordercorrleation}
\end{equation}
which in classical optics quantifies intensity correlations of a field at two space-time coordinates. Inserting Equations (\ref{eq:outputfields}) into (\ref{eq:secondordercorrleation}), we obtain sixteen terms. However, for sources with independent phase fluctuations, only six terms are non-zero \cite{Scully1997}, yielding
\begin{equation}
\begin{aligned}
G^{(2x)}(t_3,t_4)&=\frac{1}{4}|\zeta_1(t_3)\zeta_1(t_4)|^2\expect*{\hat{a}_1^{\dagger}(t_3)\hat{a}_1^{\dagger}(t_4)\hat{a}_1(t_4)\hat{a}_1(t_3)}\\
&+\frac{1}{4}|\zeta_2(t_3)\zeta_2(t_4)|^2\expect*{\hat{a}_2^{\dagger}(t_3)\hat{a}_2^{\dagger}(t_4)\hat{a}_2(t_4)\hat{a}_2(t_3)}\\
&+\frac{1}{4}|\zeta_1(t_3)|^2|\zeta_2(t_4)|^2\expect*{\hat{a}_1^{\dagger}(t_3)\hat{a}_1(t_3)\hat{a}_2^{\dagger}(t_4)\hat{a}_2(t_4)}\\
&+\frac{1}{4}|\zeta_1(t_4)|^2|\zeta_2(t_3)|^2\expect*{\hat{a}_1^{\dagger}(t_4)\hat{a}_1(t_4)\hat{a}_2^{\dagger}(t_3)\hat{a}_2(t_3)}\\
&-\frac{1}{2}\mathfrak{Re}\bigg\{\zeta_1^*(t_3)\zeta_1(t_4)\zeta_2(t_3)\zeta_2^*(t_4)\expect*{\hat{a}_1^{\dagger}(t_3)\hat{a}_1(t_4)\hat{a}_2^{\dagger}(t_4)\hat{a}_2(t_3)}\bigg\}\,.\label{eq:soccollect}
\end{aligned}
\end{equation}
Note that here we treat mode envelopes as independent of intrinsic field coherence properties. Consequently, $\zeta_k(t)$ is deterministic and can be taken out of the expectation values. This means that field modulations and statistical properties of the field are independent, which simplifies the problem and may be justified on the ground that many well-established experimental techniques for control over $\zeta_k(t)$ exist.\par
The complexity of Equation (\ref{eq:soccollect}) is greatly reduced for independent coherent states, which represent lasers. Two independent lasers are best described by a mixture of coherent states $\alpha_k=|\alpha_k|\exp(i\Theta_k)$, represented by \cite{Glauber1963,Sudarshan1963}
\begin{equation}
\hat{\rho}=\bigotimes_{k=1,2}\int_{\mathcal{C}_k}d\alpha_k\,P(\alpha_k)\ket*{\alpha_k}\bra*{\alpha_k}\,,\label{eq:cohmix}
\end{equation}
where the integration region extends over an area in the complex plane and $P(\alpha_k)$ are probability distributions. Making use of the optical equivalence theorem \cite{Sudarshan1963}, we can convert quantum-mechanical expectation values for $\hat{a}$ and $\hat{a}^{\dagger}$ into classical expectation values for coherent state (laser) amplitudes $\alpha_k$. Furthermore, the expectation values involving modes 1 and 2 factorize (due to their independence), and we obtain
\begin{equation}
\begin{aligned}
G^{(2x)}(t_3,t_4)&=\frac{1}{4}|\zeta_1(t_3)\zeta_1(t_4)|^2\bigg\langle|\alpha_1(t_3)|^2|\alpha_1(t_4)|^2\bigg\rangle_{\alpha_1}\\
&+\frac{1}{4}|\zeta_2(t_3)\zeta_2(t_4)|^2\bigg\langle|\alpha_2(t_3)|^2|\alpha_2(t_4)|^2\bigg\rangle_{\alpha_2}\\
&+\frac{1}{4}|\zeta_1(t_3)|^2|\zeta_2(t_4)|^2\bigg\langle|\alpha_1(t_3)|^2\bigg\rangle_{\alpha_1}\bigg\langle|\alpha_2(t_4)|^2\bigg\rangle_{\alpha_2}\\
&+\frac{1}{4}|\zeta_1(t_4)|^2|\zeta_2(t_3)|^2\bigg\langle|\alpha_1(t_4)|^2\bigg\rangle_{\alpha_1}\bigg\langle|\alpha_2(t_3)|^2\bigg\rangle_{\alpha_2}\\
&-\frac{1}{2}\mathfrak{Re}\bigg\{\zeta_1^*(t_3)\zeta_1(t_4)\zeta_2(t_3)\zeta_2^*(t_4)G^{(1)}_1(\tau)G^{*(1)}_2(\tau)\bigg\}\label{eq:g2correct}\,,
\end{aligned}
\end{equation}
where the expectation values are defined as ensemble averages
\begin{equation}
\expect*{f\left(\alpha_k,\alpha_k^*\right)}_{\alpha_k}=\int_{\mathcal{C}_k} d\alpha_k\,P(\alpha_k)f(\alpha_k,\alpha_k^*)\,.
\end{equation}

For CW lasers, the following three statistical assumptions are valid. First, statistical stationarity implies that the first-order autocorrelation functions
\begin{equation}
G^{(1)}_k(\tau)=\bigg\langle\alpha_k^*(t)\alpha_k(t+\tau)\bigg\rangle_{\alpha_k}\label{eq:firstorder}
\end{equation}
only depend on the detection time difference $\tau\coloneqq t_4-t_3$. Second, intensities are statistically constant and identical,
\begin{equation}
\bigg\langle|\alpha_k(t)|^2\bigg\rangle_{\alpha_k}=G^{(1)}_k(0)\equiv I_0\,.
\end{equation}
Third, intensity fluctuations are constant (for sufficiently small $\tau$),
\begin{equation}
\bigg\langle|\alpha_k(t)|^2|\alpha_k(t')|^2\bigg\rangle_{\alpha_k}=I_0^2\,.
\end{equation}
In the following, we set $I_0=1$ for convenience. In our experiment, we generate two fields from a single laser, and consequently $G^{(1)}_1(\tau)=G^{(1)}_2(\tau)\equiv G^{(1)}(\tau)$. Our laser's spectrum is best described by a Voigt line \cite{Olivero1977}, for which
\begin{equation}
G^{(1)}(\tau)=\exp\left(-\frac{|\tau|}{\tau_{\text{coh}}}-\frac{\tau^2}{\tau_{\text{coh}}^2}\right)\exp(-i\omega_0\tau)\,,\label{eq:voigtian}
\end{equation}
where $\tau_{\text{coh}}$ and $\omega_0$ are the coherence time and central frequency of the laser, respectively. The mode envelope for an unmodulated CW laser is constant, $\zeta_k(t)=1$, which gives
\begin{equation}
\begin{aligned}
G^{(2x)}(\tau)&=1-\frac{1}{2}\exp\left[-\frac{2|\tau|}{\tau_{\text{coh}}}-\frac{2\tau^2}{\tau_{\text{coh}}^2}\right]\label{eq:g2correctStationary}\,.
\end{aligned}
\end{equation}
This equation describes the Hong-Ou-Mandel dip with a visibility
\begin{equation}
V_{\text{HOM}}=\frac{G^{(2x)}(\tau)_{\text{max}}-G^{(2x)}(\tau)_{\text{min}}}{G^{(2x)}(\tau)_{\text{max}}}\label{eq:visibility}
\end{equation}
of 50\,\% for phase-randomized coherent states. It is common to measure a coincidence rate in two-photon interference experiments, which can be described by
\begin{equation}
R^{(2x)}(t_0,\tau,\Delta T)=\eta_1\eta_2\int_{t_0}^{t_0+\Delta T}dt_1\int_{t_0+\tau}^{t_0+\tau+\Delta T}dt_2\,G^{(2x)}(t_1,t_2)\,,
\end{equation}
where $\eta_1$ and $\eta_2$ denote the detection efficiencies of the single photon detectors, and $t_0$ is the initial detection time. We are interested in the time-resolved regime where the detector's time resolution $\Delta T\ll\tau_G$, where $\tau_G$ is the width of $G^{(2x)}(t_1,t_2)$, i.e. the correlation length. In this case, the measured coincidences are proportional to the fundamental quantity (correlation function), 
\begin{equation}
R^{(2x)}(t_0,\tau,\Delta T)\approx\eta_1\eta_2(\Delta T)^2G^{(2x)}(t_0,t_0+\tau)\label{eq:coinrateg2}\,,
\end{equation}
which means we can analyze $G^{(2x)}(t_1,t_2)$ in terms of the initial detection times, and thereby resolve correlation features within a time window ${\sim}\tau_G$.\par
When the mode envelopes $\zeta_k(t)$ are modulated, a structured HOM dip is obtained. Here, we are interested in polarization modulation, which leads to coincidence detection probabilities that are dependent on the optical delay $\tau_{\text{Opt}}$ between the inputs. For this we need to consider two polarization modes (designated as $H$ for horizontal polarization and $V$ for vertical) for each input mode to the HOM beam splitter. Hence, four modes with their corresponding mode envelopes $\zeta_{\sigma,k}(t)$, $\sigma\in\{H,V\}$, need to be considered. The generalization of the second-order cross-correlation function, which takes into account polarization modes, is given by
\begin{equation}
    G^{(2x)}(t_3,t_4)=G^{(2x)}_{H,H}(t_3,t_4)+G^{(2x)}_{H,V}(t_3,t_4)+G^{(2x)}_{V,H}(t_3,t_4)+G^{(2x)}_{V,V}(t_3,t_4)\,,\label{eq:g2correctpol}
\end{equation}
where the first and second polarization subscripts refer to first and second spatial modes, respectively. The correlation functions with equal polarization in both modes takes the same form as Eq. (\ref{eq:g2correct}). For the correlation functions with unequal polarization in both modes, the interference term, which is the last one in Eq. (\ref{eq:g2correct}), vanishes. Thus, from knowledge of Eq. (\ref{eq:g2correct}) we can find an expression for Eq. (\ref{eq:g2correctpol}). To do this, we first simplify Eq. (\ref{eq:g2correct}) using assumptions (\ref{eq:firstorder})-(\ref{eq:voigtian}). In addition, we consider the zero time-delay case ($\tau=0$, i.e. $t_3=t_4\equiv t_0$), for which $G^{(1)}(0)=1$. Focusing on only true coincidences allows us to isolate the dependence of the HOM interference on $\tau_{\text{Opt}}$. Using only real modulation functions, the correlation function (\ref{eq:g2correct}) gives us
\begin{equation}
\begin{aligned}
G^{(2x)}_{\sigma,\sigma}(t_0)&=\frac{1}{4}\bigg(\zeta_{\sigma,1}(t_0)^4+\zeta_{\sigma,2}(t_0)^4\bigg)\\
G^{(2x)}_{\sigma,\sigma'}(t_0)&=\frac{1}{4}\bigg(\zeta_{\sigma,1}(t_0)^2\zeta_{\sigma',1}(t_0)^2+\zeta_{\sigma,2}(t_0)^2\zeta_{\sigma',2}(t_0)^2\\&\quad+\zeta_{\sigma,1}(t_0)^2\zeta_{\sigma',2}(t_0)^2+\zeta_{\sigma,2}(t_0)^2\zeta_{\sigma',1}(t_0)^2\bigg)\label{eq:g2simplified}\,,
\end{aligned}
\end{equation}
for the equal and unequal polarization cases, respectively. We choose mode functions in the shape of square waves, which for $H$ polarization are described by 
\begin{equation}
\begin{aligned}
\zeta_{H,1}(t)&=\text{SW}_0^1\left(\frac{t-\tau_{\text{Opt}}}{T_\text{Mod}}\right)\\
\zeta_{H,2}(t)&=\text{SW}_0^1\left(\frac{t}{T_\text{Mod}}\label{eq:polmodanglesH}\right)\,,
\end{aligned}
\end{equation}
where $\text{SW}_a^b(t)$ denotes a square wave alternating between lower level $a$ and upper level $b$ (duty cycle 50\,\%, period is $T_\text{Mod}$, and $\text{SW}_a^b(0)=b$). Accordingly, the mode functions for $V$ polarization are shifted by $T_\text{Mod}/2$,
\begin{equation}
\begin{aligned}
\zeta_{V,1}(t)&=\text{SW}_0^1\left(\frac{t-T_\text{Mod}/2-\tau_{\text{Opt}}}{T_\text{Mod}}\right)\\
\zeta_{V,2}(t)&=\text{SW}_0^1\left(\frac{t-T_\text{Mod}/2}{T_\text{Mod}}\label{eq:polmodanglesV}\right)\,.
\end{aligned}
\end{equation}

Using our modulation functions (\ref{eq:polmodanglesH}) and (\ref{eq:polmodanglesV}), we arrive at a simple expression for the second-order cross-correlation function (\ref{eq:g2correctpol}),
\begin{equation}
\begin{aligned}
G^{(2x)}(t_0,\tau_{\text{Opt}})&=\frac{1}{4}\bigg(3-\text{SW}_{-1}^1\left[\frac{t_0}{T_\text{Mod}}\right]\text{SW}_{-1}^1\left[\frac{t_0-\tau_{\text{Opt}}}{T_\text{Mod}}\right]\bigg)\label{eq:G2polarizationspecmodspecial}\,.
\end{aligned}
\end{equation}
Hence, the coincidence rate can follow a square shape when analyzed with respect to the initial detection time $t_0$. The duty cycle of this HOM square wave (i.e. the duration over which coincidences are suppressed during a period $T_\text{Mod}$) depends on $\tau_{\text{Opt}}$. In particular, when $\tau_{\text{Opt}}=T_\text{Mod}/4$, the period of the coincidence rate is half of the modulator's period $T_\text{Mod}$. This frequency doubling effect of the oscillations, which is shown in Figure \ref{fig:HOMsetup} (b), is the result of the HOM interference between indistinguishable polarization states of the input beams. 

\section{Experimental realization}

\begin{figure}[t!]
\centering\includegraphics[width=\linewidth]{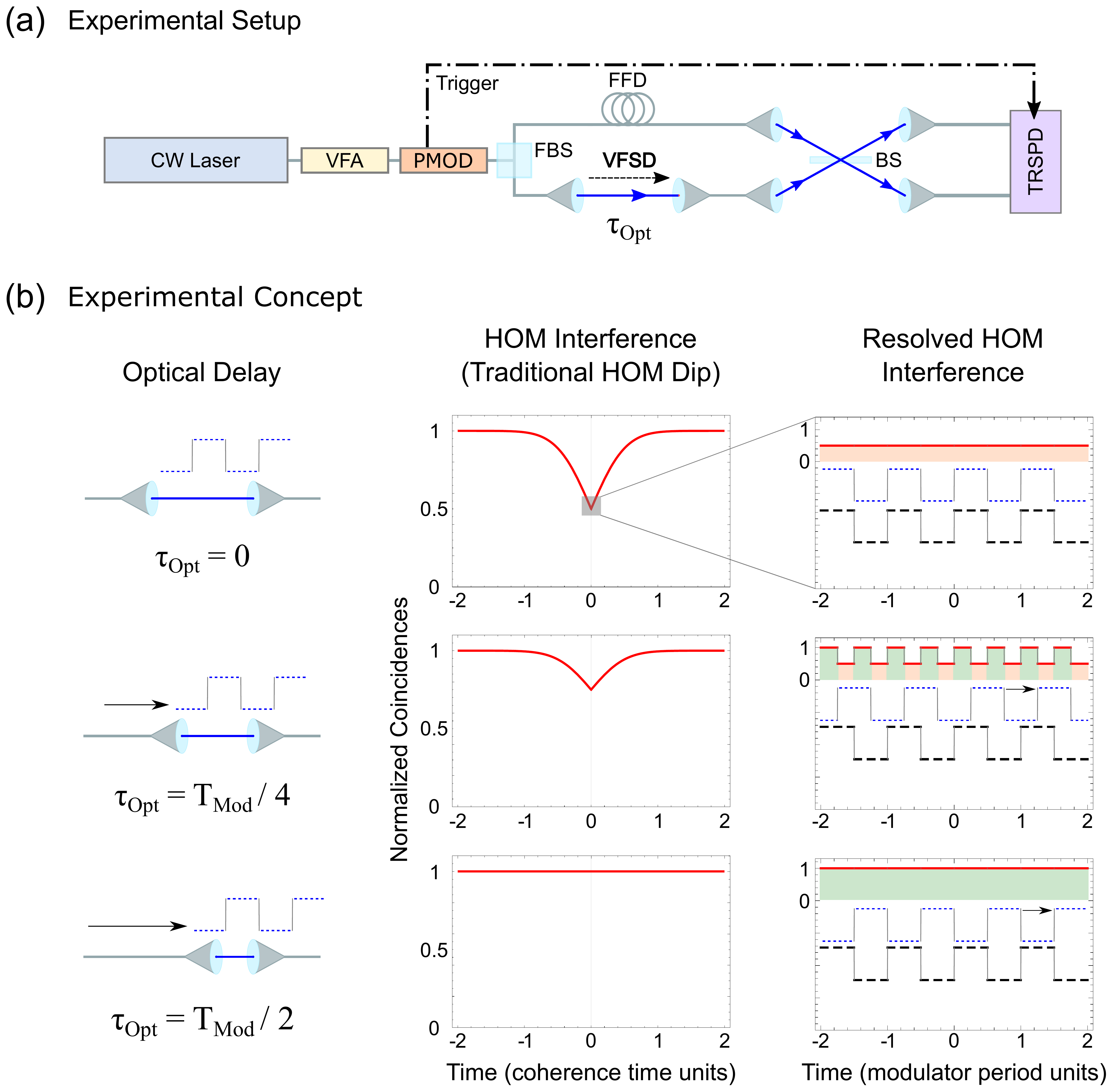}
\caption{{\bf Square HOM Waves: Experimental Setup and Concept.} {\bf(a)} A continuous-wave laser is attenuated and polarization modulated before being divided into two optical modes for the HOM interferometer. As explained in the main text, the fixed and variable time delays in the two paths are responsible for isolating the HOM interference and scanning the optical delay, respectively. {\bf(b)} Adjusting the optical delay $\tau_\text{Opt}$ changes the polarization overlap pattern between the two input modes at the HOM beam splitter (blue dotted and black dashed traces represent the two inputs), which results in changes to the two-photon coincidences (red solid traces). Y-axes for plots in (b) refer to the normalized coincidences (red solid traces) only. VFA: variable fiber attenuator, FFD: fixed 2\,km fiber delay line, VFSD: variable free-space delay line, BS: (HOM) 50:50 non-polarizing beam splitter, PMOD: polarization modulator, FBS: fiber beam splitter, TRSPD: time-resolving single-photon detectors. Note that the coherence time of the laser (microseconds) is much larger than the modulation period ($\sim$2.8\,ns).} \label{fig:HOMsetup}
\end{figure}

Figure \ref{fig:HOMsetup} (a) shows a sketch of the experimental setup. We use a CW grating-stabilized laser diode (785\,nm single mode light with $>2$\,mW intensity and microsecond-scale coherence time), and employ a variable fiber attenuator (VFA) to decrease the laser intensity down to the single-photon counting regime. The laser signal is then sent into a polarization modulator (PMOD) that acts as a fast switch between $H$ and $V$ polarization. The PMOD consists of a polarization Mach-Zehnder interferometer with a phase modulator in each path, and has been discussed elsewhere \cite{Yan2013, Pugh2017}. It is driven by an arbitrary waveform generator (AWG) at 353\,MHz, which corresponds to a modulation period of $T_\text{Mod}=2.83$\,ns. One output of the PMOD is passed through a 2\,km fixed fiber delay line (FFD) that acts as a relative dephasing channel for the two output modes because this delay is longer than the laser coherence length. The other output mode from the PMOD is sent through a variable free-space delay line (VFSD), which consists of a hollow retro-reflector moving along an optical rail. The VFSD can provide optical delays up to 3\,ns with negligible error (${\sim}10^{-3}$\,ns). The two modes are then optically combined on a 50:50 non-polarizing beam splitter (BS) in free-space (HOM beam splitter). The time-resolving single-photon detectors (TRSPD) consist of silicon avalanche photodiodes in series with time tagging units that assign time stamps to detection events with a 78.125\,ps time resolution. We also record trigger signals from the AWG that drives our PMOD.\par
Figure \ref{fig:HOMsetup} (b) illustrates the conceptual idea of how both traditional HOM dip and structured HOM interference are controlled through the relative optical delay between the two HOM input modes. Three special cases are shown. In the first case, no optical delay is imposed ($\tau_\text{Opt} = 0$), which results in varying yet identical polarization states of the two modes (blue dotted and black dashed traces represent the two inputs) at any given time that are in-phase polarization patterns. Consequently, these modes interfere perfectly at the HOM beam splitter, which reduces the normalized coincidences (red solid trace) to 0.5. In the next case, the optical delay is equal to one quarter of the modulation period ($\tau_\text{Opt} = T_\text{Mod}/4$). Thus, half the time the polarization states are identical between the input modes, and half the time they are orthogonal. Now the normalized coincidences follow a square wave pattern {\it at twice} the modulation frequency.

In the third case, the optical delay equals one half the modulation period, causing orthogonal polarization states at any given time (out-of-phase polarization patterns). Therefore, no interference ensues and the normalized coincidences trace remains at the baseline of 1. In general, for optical delays between 0 and $T_\text{Mod}/2$, the regions of overlap and non-overlap alternate. Correspondingly, the normalized coincidences oscillates between 0.5 and 1.
 
\section{Experimental results}
\begin{figure}[t!]
\centering\includegraphics[width=\linewidth]{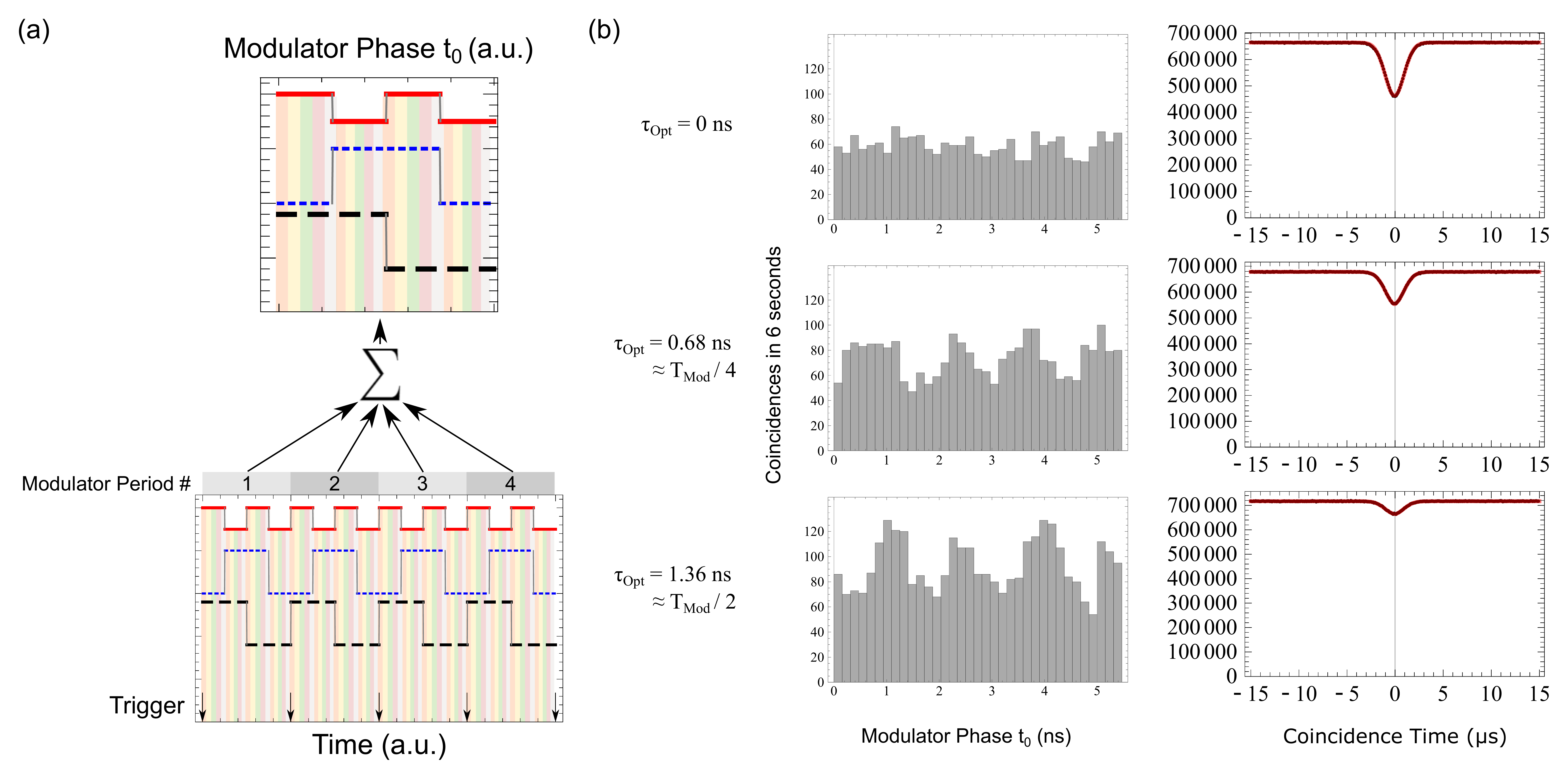}
\caption{{\bf Experimental Square Wave HOM Interference with Polarization Modulated CW Lasers.} {\bf(a)} Using time stamps of the modulation trigger as a stable time reference (indicated by the arrows in the lower plot), together with the photon detection times, we can selectively add up coincidences over time to form the resolved HOM square wave (see main text). {\bf (b)}
As expected from Equation (\ref{eq:G2polarizationspecmodspecial}), a square wave pattern with double the modulation frequency emerges for certain optical delays $\tau_{\text{Opt}}$. Coincidences are detected within a $T_{\text{coin}}=312.5$\,ps window, which approximates $\tau=0$\,ns. The histogram bin size is 156.25\,ps.} \label{fig:expsquarehom}
\end{figure}
Figure \ref{fig:expsquarehom} shows our experimental results. The mean photon number per modulation period was $3.4\cdot10^{-3}$ with a slight count rate difference between the two detectors, which contributed less than 1\,\% to the visibility loss. The measured average rate of single photon detection during a measurement time of $6$\,s for each retroreflector position was 1.2 million per second. The total number of two-photon coincidences (coincidence window $T_{\text{coin}}=312.5$\,ps) for each step was on average 4000.\par
In Figure \ref{fig:expsquarehom} (a) we illustrate how to distill the square wave predicted by Equation (\ref{eq:G2polarizationspecmodspecial}). Per modulation period, we either detect none or $\sim 1$ coincidence. To form the correlation function (\ref{eq:G2polarizationspecmodspecial}), which is a statistical quantity, we need to add up coincidences coherently over time. We utilize the modulator trigger as a stable time reference (indicated by the arrows in the lower plot): when a coincidence is found, the time-difference to the previous trigger is calculated. We refer to this time difference as the ``modulator phase'', which is the variable $t_0$ in Equation (\ref{eq:G2polarizationspecmodspecial}). Moreover, since the time resolution of our time tagging units is two orders of magnitude higher than the modulator period, we are able to resolve the square wave pattern by taking a histogram over $t_0$.\par
In Figure \ref{fig:expsquarehom} (b) we show the outcome of the analysis for optical delays corresponding to the three cases in Figure \ref{fig:HOMsetup} (b). We obtain the expected flat coincidence line for the case $\tau_{\text{Opt}}=0$\,ns . Experimentally, we identify the zero optical delay case by finding the retroreflector position in the VFSD (see Figure \ref{fig:HOMsetup} (a)) that maximizes the HOM dip visibility. We achieve a maximum visibility of $0.35\pm0.03$, which is lower than the theoretical maximum of 0.5. To assess the upper bound for the visibility in our setup, we bypass the AWG-driven polarization controller, and use manual polarization control via a half-wave plate, and achieved a maximum HOM dip visibility of $0.42\pm0.02$. Thus, due to imperfections in the AWG-driven polarization modulator and its finite rise and fall times ($\sim500$\,ps) for generating square waves, we incur a $7$\,\% visibility drop.\par
Next, we move the retroreflector to a position corresponding to a 0.68\,ns optical delay, which approximates $T_{\text{Mod}}/4\approx0.7$\,ns. We now observe a square wave of coincidences while simultaneously the HOM dip visibility reduces to roughly half the maximum ($\sim18$\,\%). This reduction is because half the time the coincident photons have orthogonal polarization within the dip. Finally, we set the delay to 1.36\,ns, corresponding to $T_{\text{Mod}}/2\approx1.4$\,ns. We can clearly see that the coincidence counts now oscillate between 70 and 120 coincidences. An explanation for this modulation pattern is given in Figure \ref{fig:nonidealdutycycle}, and is discussed in detail in the next section. While one may have expected a flat line corresponding to maximum coincidences, two experimental effects cause the resolved HOM interference to modulate.\par
\begin{figure}[ht!]
\centering\includegraphics[width=\linewidth]{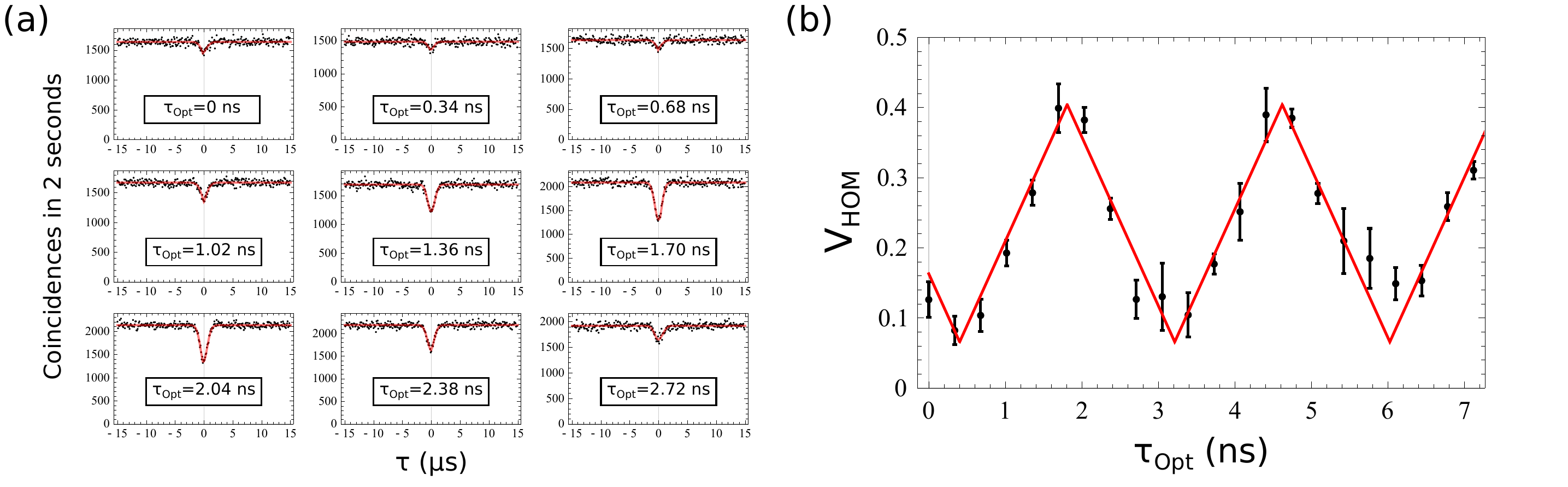}
\caption{{\bf Experimental Triangle Wave HOM Interference with CW Lasers.} {\bf(a)} Traditional HOM dips extracted for optical delays (retroreflector positions) ranging over one modulator period $T_{\text{Mod}}$. Black dots: data points, red solid line: fit. {\bf(b)} Plot of HOM dip visibilities for optical delays ranging from $0$ to $7.14$\,ns. The red curve is a fit described by Equation (\ref{eq:G2triangle2}) for $\tau\approx0$. The period of the fitted waveform is $T=(2.80\pm0.04$)\,ns, which matches the modulator period as expected. Compared with the measurements presented in Fig. \ref{fig:expsquarehom}, here our setup had an offset of the optical delay of (0.85$\pm$0.02)\,ns, which the fit takes into account.} \label{fig:fitfigure}
\end{figure}

Our setup can also be used to generate triangle HOM waves. If we ignore the modulator phase, and thus integrate (\ref{eq:G2polarizationspecmodspecial}) over $t_0$ for a measurement time $T_{\text{M}}$, we obtain 
\begin{equation}
G^{(2x)}(\tau_{\text{Opt}})=T_{\text{M}}\bigg(1-\frac{1}{2}\text{TW}_0^1\left[\frac{\tau_{\text{Opt}}}{T_\text{Mod}}-\frac{3}{4}\right]\bigg)\,,\label{eq:G2triangle2}
\end{equation}
where $\text{TW}_a^b(t)$ is a triangle wave with minimum $a$, maximum $b$, and $\text{TW}_a^b(0)=0$. Thus, without a phase reference from the modulation trigger, the correlation function follows a triangular shape with the same period as the modulator, $T_\text{Mod}$. Note that this is true for $\tau\approx 0$ as before (i.e. true coincidences). We carried out another measurement, using the following simple way to extract the triangle wave. In Figure \ref{fig:fitfigure} (a) we show the HOM dip measurement for various optical delays, as described by Equation (\ref{eq:g2correctStationary}). Note that the normalized coincidences at $\tau=0$ are identical to the dip depth. Furthermore, a plot of the visibility Eq. (\ref{eq:visibility}) as a function of optical delay $\tau_{\text{Opt}}$ represents a triangle function of optical delay, as expected from Eq. (\ref{eq:G2triangle2}). Our results plotted in Figure \ref{fig:fitfigure} confirm this is indeed the case. The visibility does not reach zero because the optical delay is not matched for perfect anti-overlap of the square waves. Also, the finite rise and fall times of the polarization modulator in conjunction with a non-symmetric duty cycle reduces the visibility. 
\section{Discussion}
The traditional HOM dip is a feature of the full HOM interference pattern that appears on the coherence length time scale of the fields. Traces of high-speed field modulations are present in HOM interference, but are not discernible at the coherence length time scale. We have shown this in our experiment for square wave modulation, where the ratio of modulation period to coherence length is $\sim10^{-9}/10^{-6}=10^{-3}$. Crucially, though, any other modulation function would have resulted in identical HOM dips, which is determined entirely by the unmodulated field's autocorrelation function, or spectrum. In other words, infinitely many ``HOM waves'' give rise to the same HOM dip, and alleviation of this uncertainty (and revelation of hidden information) required, in our case, single-photon detectors with sub-nanosecond time-resolution.\par
\begin{figure}[ht!]
\centering\includegraphics[width=\linewidth]{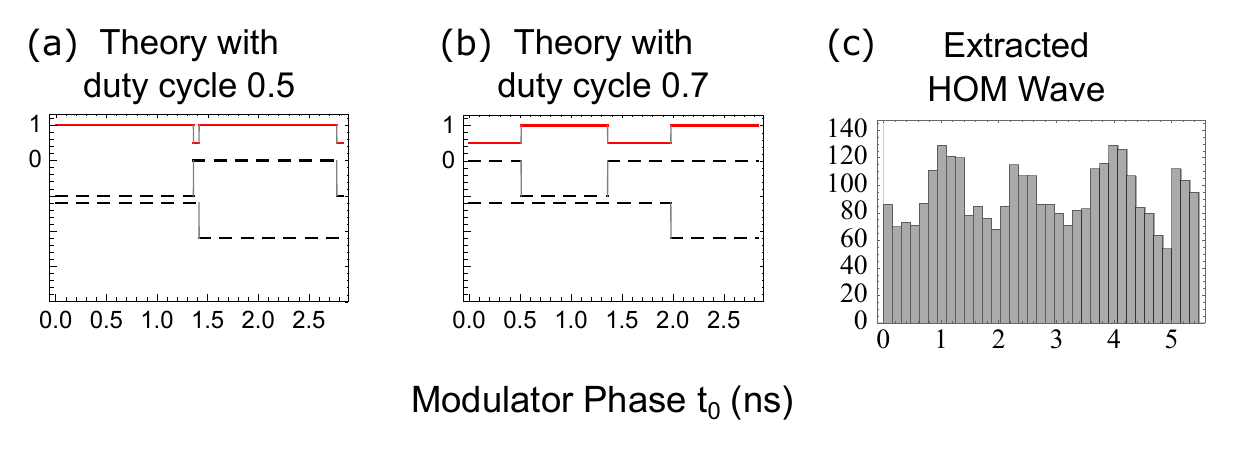}
\caption{{\bf Effect of non-ideal optical delay and asymmetrical duty cycle settings.} (a) and (b) are theory plots showing the predicted normalized coincidences (red solid line) and the polarization patterns (black dashed lines). (a) Non-ideal (offset) optical delay with a symmetrical 0.5 duty cycle, (b) non-ideal optical delay and an asymmetrical duty cycle of 0.7 are taken into account in Eq. (\ref{eq:G2polarizationspecmodspecial}) for the $\tau_{\text{Opt}}=T_{\text{Mod}}/2$ case. As a consequence, the extracted experimental HOM interference pattern (c) does not show a flat line at maximum coincidences. Note that for clarity, the x-axis for (a) and (b) graphs extend only over one modulator period, whereas the data in (c) extends over two modulator periods.} \label{fig:nonidealdutycycle}
\end{figure}

There are two effects that currently limit the performance. The first effect arises from the non-symmetric duty cycle of our polarization modulator. It deviates substantially from the symmetric value of 50\,\%, in which case we should get, for instance, $H$-polarized photons during the first half of the modulation period, and $V$-polarized photons the other half. However, a polarization modulator with a duty cycle of 70\,\% produces, for instance, $H$-polarized photons during nearly $3/4$ of the modulation period. The theory plot for the case where the duty cycle equals 0.7 in Figure \ref{fig:nonidealdutycycle} shows the polarization patterns (black dashed lines) and resulting coincidences (solid red line). We see that, even though the optical delay is set correctly to half the modulation period, the coincidences do not follow a flat line because the two polarization patterns are not complementary. This duty cycle effect does not affect the case where the optical delay is zero because non-shifted polarization patterns overlap perfectly no matter what their shape. This is why we still obtain a flat line at minimum coincidences in Figure \ref{fig:expsquarehom} (b). A second effect arises when the optical delay does not match precisely, for example, $T_{\text{Mod}}/2$. Then the polarization patterns also do not perfectly overlap, as the theory plot for a duty cycle of 50\,\% shows, though the effect is much smaller than that for the asymmetric duty cycle. Hence, if Eq. (\ref{eq:G2polarizationspecmodspecial}) is used with both a non-ideal optical delay and asymmetric duty cycle, the detected coincidences, and thus the HOM square wave, follows the predicted shape. The theory could also be expanded to include a varying input intensity to explore how it may affect the coincidence signal but this is beyond the scope of our present work.
\section{Conclusion}
In summary, we have modelled and observed Hong-Ou-Mandel square and triangle wave interference patterns using polarization-modulated continuous-wave lasers. The basic equation is (\ref{eq:g2correct}), which shows that any kind of laser field modulation affecting the mode envelopes $\zeta_k(t)$ will result in visible HOM interference patterns. Our work demonstrates a striking feature of Hong-Ou-Mandel interference between coherent states: despite different envelope functions $\zeta_k(t)$, the full interference visibility can still be obtained within the region of temporal overlap. As in our case, we experimentally recovered clear HOM interference between wavepackets that are identical in shape but shifted in time. This seems counterintuitive, as traditional HOM experiments are based on well-defined wavepackets and careful path alignment to ensure they overlap at the interfering beam splitter.\par
Our results can apply to various modulation schemes to encode information. For example, a sender could encode information in the polarization state of photons, and a receiver could use HOM dip visibility levels to decode the message. This will be particularly useful in cases where alignment of the reference frames is difficult or not possible (communication without a shared Cartesian frame) \cite{Bartlett2007,Tannous2019}. Our findings could be applicable to HOM-based quantum protocols which are sensitive to temporal alignment where fluctuating optical path lengths can degrade the HOM effect. In particular, free-space quantum communication links with moving platforms, such as between vehicles, aircraft and/or satellites \cite{Bourgoin2015,Pugh2017}, may have rapidly changing path lengths that are difficult to compensate for in real time. For instance, the measurement-device independent quantum key distribution implementation of Yuan {\it et al.} \cite{Yuan2014} had a $50$\,ps misalignment that prevented HOM interference. Our work is a proof-of-concept system which, after improvements to the polarization modulator and using independent lasers, could be utilized for such protocols.

\section*{Funding}
This research was supported in part by the Office of Naval Research; Canadian Space Agency; Canada Foundation for Innovation (25403, 30833); Ontario Research Foundation (098, RE08-051); Canadian Institute for Advanced Research; Natural Sciences and Engineering Research Council of Canada (RGPIN-386329-2010); Industry Canada.
\section*{Acknowledgments}
The authors would like to thank Matteo Mariantoni for lending us the high-speed arbitrary waveform generator.
\section*{Disclosures}
The authors declare no conflicts of interest.
\newline
\bibliographystyle{IEEEtran}
\bibliography{laserhomdip}

\end{document}